\documentclass[12pt]{article}
\usepackage{axodraw}
\usepackage{amsmath}
\usepackage{graphicx}

\def\be{\begin{equation}}
\def\ee{\end{equation}}
\def\bea{\begin{eqnarray}}
\def\eea{\end{eqnarray}}

\def\e{\epsilon}

\def\m{\mu}
\def\n{\nu}

\def\lsim{\raise0.3ex\hbox{$\;<$\kern-0.75em\raise-1.1ex\hbox{$\sim\;$}}}
\def\gsim{\raise0.3ex\hbox{$\;>$\kern-0.75em\raise-1.1ex\hbox{$\sim\;$}}}

\def\inbar{\,\vrule height1.5ex width.4pt depth0pt}

\def\IC{\relax\hbox{$\inbar\kern-.3em{\rm C}$}}
\def\IQ{\relax\hbox{$\inbar\kern-.3em{\rm Q}$}}
\def\IR{\relax{\rm I\kern-.18em R}}
 \font\cmss=cmss10 \font\cmsss=cmss10 at 7pt
\def\IZ{\relax\ifmmode\mathchoice
 {\hbox{\cmss Z\kern-.4em Z}}{\hbox{\cmss Z\kern-.4em Z}}
 {\lower.9pt\hbox{\cmsss Z\kern-.4em Z}}
 {\lower1.2pt\hbox{\cmsss Z\kern-.4em Z}}\else{\cmss Z\kern-.4em Z}\fi}

\begin{document}

\vspace{-1truecm}

\rightline{CPHT RR-024.0409}
\rightline{LPT--Orsay 09/26}
%\rightline{\today}

\vspace{0.cm}

\begin{center}

{\Large {\bf (In)visible $Z'$ and dark matter }}
\vspace{1 cm}\\

{\large E. Dudas$^{1,2}$, \ Y. Mambrini$^2$, S. Pokorski$^{3}$ and \ A. Romagnoni$^{2,1}$
}
\vspace{1cm}\\

$^1$
CPhT, Ecole Polytechnique 91128 Palaiseau Cedex, France
\vspace{0.3cm}\\

$^2$
Laboratoire de Physique Th\'eorique,
Universit\'e Paris-Sud, F-91405 Orsay, France
\vspace{0.3cm}\\

$^3$
Institute of Theoretical Physics, Warsaw University, Hoza 69, 00-681 Warsaw,
Poland
\vspace{0.3cm}\\

\end{center}

\vspace{1cm}

\abstract{\noindent
We study the consequences of an extension of the standard model containing
an invisible extra gauge group under which the SM particles are neutral. We show
that effective operators,  generated by loops of heavy chiral fermions charged under
both gauge groups and connecting the new gauge sector to the Standard Model, can
give rise to a viable dark matter candidate. Its annihilations produce clean visible signals through a gamma-ray line.
This would be a smoking gun signature of such models observable by actual experiments.
}

\newpage

\vspace{3cm}

\newpage

%\tableofcontents

\vspace{3cm}

%\newpage

%\setcounter{page}{1}
\pagestyle{plain}

\section{Introduction}

Extensions of the SM with additional Z' symmetries were widely
discussed in the literature \cite{zprime}.
There are several ways of defining semi-invisible $Z'$ theories, under
which SM fermions are neutral, with effective operators connecting
directly $Z'$ to the SM sector. The simplest one can imagine
is connecting the two sectors by a kinetic mixing operator
\begin{equation}
\delta \cdot \ F_X^{\mu \nu} \ F^Y_{\mu \nu} \ ,  \label{i1}
\end{equation}
with $F^I_{\rho \sigma}= \partial_{\rho} B^I_{\sigma} - \partial_{\sigma} B^I_{\rho}$,
which is naturally generated at the loop-level by heavy fermions
charged both under SM and the $Z'$ gauge symmetry.
This scenario is very interesting from a dark matter perspective and
was studied intensively over the last years
\cite{zmixing,Feldman:2007wj,Ring}.
The second and less studied is the effect of the vertex connecting
$Z'$ to two SM gauge fields.  The simplest
examples  are the generalized Chern-Simons (GCS) terms \cite{abdk,gcs,ferrara}, for example
\begin{equation}
\epsilon^{\mu \nu \rho \sigma} Z'_{\mu} B_{\nu} F_{\rho \sigma}^Y \ , \label{i2}
\end{equation}
which were argued to be generated by heavy fermions in \cite{abdk}.

In this letter we study from an effective operator viewpoint the GCS
terms of the type (\ref{i2}) and their decoupling properties
at low-energy. We argue that the $Z' Z \gamma$ and $Z' W W $  vertices
necessarily have a heavy-fermion mass suppression $1/M^2$. This is
due to the fact that, if the heavy fermion masses $M$ which are
decoupled ($M \rightarrow \infty$) are SM symmetric, coming only from
the Higgs mechanism breaking the $Z'$ gauge symmetry, the operator
(\ref{i2}) should be invariant under the non-linearly (in the broken-phase)
realized $Z'$ symmetry, whereas it should be well-defined (non-singular)
in the unbroken SM phase. The gauge invariant version of  (\ref{i2})
is then
\begin{equation}
\frac{i}{M^2} \ \epsilon^{\mu \nu \rho \sigma} {\cal D}_{\mu} \theta_X ( H^+ {D}_{\nu} H
-  ({D}_{\nu} H)^+ H ) \ F_{\rho \sigma}^Y \label{i3}
\end{equation}
where
 $D_{\nu}$ is the generic covariant derivative, ${\cal D}_{\mu}$
 defined as  ${\cal D}_{\mu} \theta_X = \partial_{\mu} \theta_X - g_X
Z'_{\mu}$ is the Stueckelberg gauge-invariant combination
with $\theta_X$ being the Stueckelberg axion, $g_X$ is the $Z'$ gauge coupling and
$M$ is related to the heavy fermion masses.
After electroweak symmetry breaking, we get a GCS term (\ref{i2}) with
a coeff. prop. to  $v^2/M^2$, where $v$ is the electroweak vev.

In the case of two $Z'$ gauge bosons, we show that there is a genuine
non-decoupling effect, i.e. independent of the heavy-fermion masses\footnote{see
  \cite{aw} for a recent similar example.}.
Indeed, for two extra gauge symmetries $U(1)_X$ and $U(1)_X'$ with
Stueckelberg realization of gauge symmetries , the dimension-four operator
\begin{equation}
\epsilon^{\mu \nu \rho \sigma} {\cal D}_{\mu} \theta_X  {\cal D}_{\nu}
\theta_X' F_{\rho \sigma}^Y \ ,
\label{tree}
\end{equation}
where $\theta_X$ and $\theta_X'$ are the axions of the two Z'
respectively, is gauge invariant and can be generated by a heavy chiral but
anomaly-free fermion spectrum charged under the two $U(1)$'s and the
SM. We will check this explicitly in Section 4  by using the formulae of
ref. \cite{abdk}.
The masses of the heavy fermions that are taken to infinity $M
\rightarrow \infty$  have to come from the Higgs breaking of the two
$U(1)$'s and has to be SM invariant.
In both cases, of one or two $Z'$, we allow possible SM-like mass
elements $m \sim v$ for the heavy fermions, which we keep fixed in the
decoupling limit $M \rightarrow \infty$.
 The term  (\ref{tree}) provides
an interesting counter-example of the decoupling theorem
\cite{appelquist}. This is different from the non-decoupling effects studied
in \cite{dhoker} for two reasons. First, the heavy fermionic spectrum
we will consider, albeit chiral, is free of any gauge
anomalies. Secondly, whereas in \cite{dhoker} the fermions which were
decoupled had a SM-like mass, in the case studied here the
masses which decouple are SM invariant.

The GCS terms were already discussed from the viewpoint of anomalous three-gauge boson
vertices, notably $Z' Z \gamma$ in various papers
\cite{abdk,coriano,wells}. The consequences for the LHC were
subsequently  analyzed in \cite{coriano,wells,gianfranco} and
\cite{aw}.  In ref. \cite{fucito} the supersymmetric partner of the
axion, the axino, was discussed as a dark matter candidate .
The main point of the present paper is that, whereas the mass
suppression in (\ref{i3}) make the LHC signatures of such an
(in)visible $Z'$ difficult to detect, the interactions described by
the dimension-six operators (\ref{i3}) and others similar to it,
discussed in more detail in Section 2, make
the lightest fermion in the $Z'$ sector a viable dark matter
candidate. Indeed, due to the couplings
(\ref{i2}) such a fermion can annihilate into a $Z$ and a photon, via
the s-channel $Z'$ virtual exchange, with an appropriate relic
abundance. An interesting signature of this
channel of dark-matter annihilation is the gamma ray in the final
state, which is monochromatic and can be tested with the  FERMI/GLAST \cite{Gehrels:1999ri}
experiment in the near future. We would like to emphasize that the reason this
(loop-suppressed) coupling can produce a visible gamma ray signal is
that in our case it is the same diagram which describes the main annihilation
channel for the dark matter and simultaneously generate the
monochromatic gamma ray. In contrast, in other BSM models (e.g
supersymmetry), the $Z'Z \gamma$ vertex is loop suppressed, whereas
the dark matter annihilation occurs at tree-level, making the gamma
ray signal highly suppressed.
We argue that if the vertex (\ref{i2}) dominates\footnote{This can be
  realized if the heavy fermions generating these operators come in
  complete $SU(5)$ representations.} or gives similar
effects compared to the kinetic mixing (\ref{i1}), the monochromatic
gamma ray remains visible, providing an astrophysical window towards high-energy physics.
   For the two $Z'$ case, the unsuppressed coupling  (\ref{tree}) leads
to an enhancement of the diagram producing the relic density and the
monochromatic gamma ray, which now is generated by the anomalous
coupling $Z' Z^{''} \gamma$, provided the kinematic constraint $M_{Z"} < 2 M_{DM}$ is fulfilled. For TeV scale DM mass, both
$Z'$ and $Z"$ should then have TeV masses, in order to get good relic density via the DM annihilation into $ Z^{''} \gamma$.

%%%%%%%%%%%%%%%%%%%%%%%%%%%%%%%%%%%%%%%%%%%%%%%%%%%%%%%%%%%%%%%%%%%%%%%%%%%%%%%%%%

\section{(In)visible $Z'$, effective operators and decoupling}
\label{section1}

We consider here an effective model where a left and right dark matter
fermion $\psi^{DM}_L$, $\psi^{DM}_R$,  charged under a spontaneously broken
extra $U(1)_X$, with charges $X^{DM}_L$ and $X^{DM}_R$ respectively, is added to the Standard
Model sector. If the $U(1)_X$ is invisible to the SM, i.e if quarks
and leptons are neutral with respect to the extra gauge symmetry,
the only way $Z'$ can contribute to the low-energy physics, is through
effective interactions obtained after integrating out the UV physics.
Typical examples of this kind of effects are given by considering a
heavy sector of fermions charged under $U(1)_X$ and the
SM gauge group, that we briefly discuss later on. When the heavy fermions decouple, loop effects give rise
%, as we will discuss more explicitly in Section \ref{Examples}.\\
to the general effective Lagrangian:
\begin{eqnarray}
&& {\cal L} \ = \ {\cal L}_{SM} + \  {\bar \psi}_L^{DM} \left( i \gamma^{\m}
\partial_{\m} + g_X X^{DM}_L \gamma^{\m} Z'_{\mu} \right)
\psi^{DM}_L + {\bar \psi}^{DM}_R \left( i \gamma^{\m}
\partial_{\m} + g_X X^{DM}_R Z'_{\mu} \right)
\psi^{DM}_R \nonumber \\
&& - \left( {\bar \psi}^{DM}_L M_{DM} \psi^{DM}_R
+ {\rm h.c.} \right) \  + \ {1 \over 2} (\partial_{\mu} a_X - M_{Z'} Z'_{\mu} )^{2} -
\frac{1}{4} F^X_{\mu \nu} F^{X \, \mu \nu} \nonumber \\
&& + {\cal L}_{1}(Z'_{\mu}) + {\cal L}_{2}(B_{\mu}, W^a_{\mu}) + {\cal L}_{mix}(Z'_{\mu}, B_{\mu}, W^a_{\mu}) \ ,
%+  \alpha \ \epsilon^{\mu \nu \rho \sigma} ( Z'_{\mu} - {1 \over
%M_{Z'}} \partial_{\mu} a ) B_{\nu} B_{\rho \sigma} \ ,
\label{inv1}
\end{eqnarray}
where ${\cal L}_{SM}$ is the Standard Model Lagrangian, ${\cal L}_{1}$
and ${\cal L}_{2}$ represent the new effective operators
 generated separately in the SM gauge sector and $Z'$ one, whereas in
 ${\cal L}_{mix}$ we collect all the induced terms mixing them. The
 aim of this paper is to show how these terms give the possibility to
 detect the presene of an "invisible" $Z'$.
$Z'$ gauge symmetry is spontaneously broken by the vev of a Higgs field $S$.
The Stueckelberg axion $a_X$ assures the gauge invariance of the
effective action, and $g_X$ and $ F^X$
are the $Z'$ gauge coupling and gauge field strength. The Stueckelberg
mechanism can be understood as a heavy Higgs mechanism, where the
extra Higgs field $S$ takes the
form $S= (V + s) ~\exp[i \frac{a_X}{V}]$, where $V$ is the heavy Higgs
vev, and the axion transforms non-linearly under $U(1)_X$ gauge transformations
\begin{equation}
\delta A^{\mu}_X \ = \ \partial^{\mu} \alpha \quad , \quad \delta a_X
\ = \ \alpha \  g_X V \ ,
\end{equation}
for a Higgs field $S$ of $X$-charge equal to $1$.

The lightest fermion charged (only) under Z' will be our dark matter
candidate.  The dark matter mass $M_{DM}$ can be of two types~: \\
- If $\Psi^{DM}_L$ and $\Psi^{DM}_R$ have equal charges  $X^{DM}_L =
X^{DM}_R$ , then $\Psi^{DM}$ is vector-like and therefore we can write the  Dirac mass $M {\bar
\Psi}^{DM} \Psi^{DM}$. The magnitude of the DM mass in this case is completely unrelated to the $Z'$ mass. \\
- If they are chiral, i.e. the left and right $U(1)_X$ charges are
different $X^{DM}_L = X^{DM}_R  \pm 1$, then we can write down  Higgs-type
masses \ $\lambda_{DM}  S {\bar \Psi}^{DM}_{L} \Psi^{DM}_R + \ {\rm h.c.}$
\ or
$\lambda_{DM}  S^+ {\bar \Psi}^{DM}_L \Psi^{DM}_R$  $+ {\rm h.c}$. In this case, for DM Yukawa couplings of the order of the
$Z'$ gauge couplings, naturally $M_{DM} \sim M_{Z'}$.  \\
   Since we will be interested in electroweak values for DM mass, in both cases we will consider a standard range
$ 100 \ GeV \lsim  M_{DM} \lsim 1 \ TeV$. \\
The Higgs $S$ can be also invoked to provide a mass for
the heavy fermions.

\noindent
Let us enter more into the details of the effective interactions. For notations convenience we define:
\begin{eqnarray}
 \theta_X  \equiv \frac{a_X}{V} \quad , \quad && ~{\cal D}_{\mu}
 \theta_X \equiv \ \partial_{\mu} \theta_X  - g_X Z'_{\mu} \ , \nonumber \\
%{\cal D}_{\mu} \theta_i &\equiv& \left(\partial_{\mu} \theta_i  - g' Z'_{\mu} \right) \nonumber \\
\widetilde{F}_{\mu \nu} \equiv \epsilon_{\mu \nu \rho \sigma} F^{\rho \sigma} \quad ,  &&(F G) \equiv {\rm Tr} [ F_{\mu \nu} G^{\mu \nu} ]
\quad  , ~~{\cal T}r (E F G) \equiv {\rm Tr} [ E_{\mu}^{~\lambda} F_{\lambda \nu} G^{\nu \mu}] \ ,
\end{eqnarray}
where ${\rm Tr}$ takes into account a possible trace over non-abelian indices.
Of crucial importance in what follows are the symmetries of the high-energy theory, which includes the heavy fermions $\Psi^{(H)}_{L,R}$
transforming under both $U(1)_X$ and the SM gauge group. There are two cases :
\begin{itemize}

\item If they are vector-like, i.e. the left and right $U(1)_X$ charges are
equal $X_L = X_R$ and therefore they have Dirac masses $M {\bar \Psi}^{(H)}_L \Psi^{(H)}_R + {\rm h.c.}$, the effective operators
obtained after integrating them out have to respect the charge conjugation symmetry $C$. In this case, a straightforward generalization
of the Furry's theorem applies and the first effective operator constructed out of gauge fields mixing the two sectors
is of the Euler-Heisenberg type $(1/ M^4) F^4$.  Due to the big $M^4$ mass suppression, this is not the case of interest for us.

\item   If they are chiral, i.e. the left and right $U(1)_X$ charges are
different $X_L = X_R \pm 1$ and therefore they have Higgs-type masses $\lambda  S {\bar \Psi}^{(H)}_L \Psi^{(H)}_R + {\rm h.c.}$ or
$\lambda  S^+ {\bar \Psi}^{(H)}_L \Psi^{(H)}_R + {\rm h.c.}$ , the effective operators
obtained after integrating them out violate $C$.  They respect the $CP$ symmetry if all couplings are real and can violate
CP for complex couplings.
\end{itemize}
We use gauge invariance and CP symmetry of the lagrangian in order to classify
the effective interaction terms invariant under $SU(2) \times U_Y(1)
\times U(1)_X$ at low-energy. An important point in what follows is
that, while the $U(1)_X$ gauge symmetry is necessarily realized
in the broken (Stueckelberg) phase, if they are generated by heavy
states respecting the SM gauge symmetry,  the effective operators have
to be invariant under the unbroken SM gauge group. CP symmetry is a
useful tool in classifying the effective operators
since non-decoupling (mass-independent) effects have to
respect it. We restrict in what follows for simplicity to CP-even operators.\footnote{The CP-odd operators have the form:
\item{Dimension-four, CP-odd operators : $ {\cal D}_{\mu} \theta_X
(H^{\dagger} {D}^{\mu} H + c.c.),  \quad \partial^{\m} {\cal D}_{\mu} \theta_X H^{\dagger} H$ . }
%\begin{equation}
%\delta_3 \ {\cal D}_{\mu} \theta_X (H^{\dagger} {\cal D}_{\mu} H + {\cal D}_{\mu} H^{\dagger} H) \quad , \quad
%\delta_4 \partial^{\m} {\cal D}_{\mu} \theta_X H^{\dagger} H \ .
%\end{equation}
\item{Dimension-six, CP-odd: $ \frac{1}{M^2}  {\cal D}^{\mu} \theta_X
\left[ i ({D}^{\nu} H)^{\dag}  F^V_{\mu \nu} H + c.c. \right],   \quad
\frac{1}{M^2}    \left[ ({D}^{\nu} H)^{\dag} \widetilde{F}^V_{\mu \nu} H + c.c. \right]$ \ , \\
\hspace*{4.4cm}$\frac{1}{M^2} \partial^{\mu} {\cal D}_{\mu} \theta_X (F^V F^V), \quad
 \frac{1}{M^2} {\cal D}_{\mu} \theta_X {\cal D}^{\mu} \theta_X (F^{V} \widetilde{F}^{V})
$ \ .} \\
We remind the reader that the axion $a_X$ is CP odd.
Some of these operators could be of some interest for CP violation in the Higgs sector, but this is beyond the goals of the
present paper.}
Restricting then to CP-invariant operators mixing the two sectors, we then find~:
\begin{itemize}
\item Dimension-four operators~ :
\begin{equation}
\delta \ F^Y_{\mu \nu} F^{X \, \mu \nu} \qquad , \qquad i \eta
\ {\cal D}_{\mu} \theta_X H^{\dagger} { D}_{\mu} H + c.c.
%- {\cal D}_{\mu} H^{\dagger} H) \ .
\label{l02}
\end{equation}
%\item Dimension-six operators with no mixing between Z' and SM~:
%\begin{equation}
%\frac{1}{M^2} ~\widetilde{d}_1 ~ \partial^{\mu} {\cal D}_{\mu} \theta_X \ (F^X \widetilde{F}^X) \qquad , \qquad
%\frac{1}{M^2}~ \widetilde{b}_1 ~ {\cal T}r (F^Y F^W \widetilde{F}^W)  \ \label{l2}
%\end{equation}
\item Dimension-six operators :
\begin{eqnarray}
&&{\cal L}_{mix} \ = \ \frac{1}{M^2}
\Big\{  b_1 {\cal T}r (F^X F^Y \widetilde{F}^Y) +  2 b_2 {\cal T}r (F^X F^W \widetilde{F}^W) +  b_3 {\cal T}r (F^Y F^X \widetilde{F}^X)  \nonumber \\
&& \quad + {\cal D}^{\mu} \theta_X
\left[ i ({D}^{\nu} H)^{\dag} (c_1  \widetilde{F}^Y_{\mu \nu} + c_2  \widetilde{F}^W_{\mu \nu}
 + c_3  \widetilde{F}^X_{\mu \nu}) H + c.c.
% - H^{\dag} (c_1 \widetilde{F}^Y_{\mu \nu} + c_2  \widetilde{F}^W_{\mu \nu}+ c_3  \widetilde{F}^X_{\mu \nu}) {\cal D}^{\nu} H
\right] \nonumber \\
%%&& \quad +  {\cal D}^{\mu} \theta_X
%%\left[ ({D}^{\nu} H)^{\dag} (c_4 F^Y_{\mu \nu} + c_5 F^W_{\mu \nu} +
%%  c_6 F^X_{\mu \nu}) H + c.c.
% H^{\dag} (c_4 F^Y_{\mu \nu} + c_5 F^W_{\mu \nu}+ c_6 F^X_{\mu \nu}) {\cal D}^{\nu} H
%%\right]  \nonumber \\
&&  \quad + \partial^{\mu} {\cal D}_{\mu} \theta_X \Big[ d_1 (F^Y
\widetilde{F}^Y) + 2 d_2 (F^W \widetilde{F}^W) + d_3 (F^Y
\widetilde{F}^X) \Big]  \nonumber \\
&& \quad + {\cal D}_{\mu} \theta_X {\cal D}^{\mu} \theta_X \ \Big[ d_4
(F^{Y} F^{Y}) + 2 d_5 (F^{W} F^{W}) \Big] \Big\}  \ . \label{lmix}
\end{eqnarray}
\end{itemize}

%\begin{eqnarray}
%&& \frac{1}{M^2}  {\cal D}^{\mu} \theta_X
%\left[ i ({\cal D}^{\nu} H)^{\dag} (c'_1 F^Y_{\mu \nu} + c'_2 F^W_{\mu \nu} +
%  c'_3 F^X_{\mu \nu}) H + c.c.
%%- H^{\dag} (c'_1 F^Y_{\mu \nu} + c'_2 F^W_{\mu \nu}+ c'_3 F^X_{\mu \nu}) {\cal D}^{\nu} H
%\right] \ , \nonumber \\
%&& \frac{1}{M^2}    \left[ ({\cal D}^{\nu} H)^{\dag} (c'_4  \widetilde{F}^Y_{\mu \nu} + c'_5  \widetilde{F}^W_{\mu \nu} + c'_6 \widetilde{F}^X_{\mu \nu}) H + c.c.
%%H^{\dag} (c'_4 \widetilde{F}^Y_{\mu \nu} + c'_5  \widetilde{F}^W_{\mu \nu}+ c'_6  \widetilde{F}^X_{\mu \nu}) {\cal D}^{\nu} H
%\right] \nonumber \\

%
%&& \frac{1}{M^2} \partial^{\mu} {\cal D}_{\mu} \theta_X \Big[ d'_1 (F^Y F^Y) +   d'_2
%(F^W F^W)  + d'_3 (F^Y F^X)+ \widetilde{d}'_1 (F^X F^X) \Big] \ , \nonumber \\
%&& \frac{1}{M^2} {\cal D}_{\mu} \theta_X {\cal D}^{\mu} \theta_X \ \Big[d'_4 (F^{Y} \widetilde{F}^{Y}) +  d'_5 (F^{W} \widetilde{F}^{W}) \Big]
%\end{eqnarray}
%\end{itemize}

\noindent
%and $Z'_{\mu}$ takes mass by the Stuckelberg mechanism, due to the coupling with an extra heavy boson, $S$, charged only under  $U(1)_X$.\\
%Following the  usual notation, $D_{\nu}=\partial_\nu - i
%\sum_k g_k x_k B_{\nu}^k $ is the
%covariant derivative (with $x_k$ the charge with respect to the $k$-th group, and $B_{\nu}^k$ the related gauge field)
% and the field strength are defined as:
% % of the $k$-th gauge fields:
% $F^Y_{\rho \sigma}= \partial_{\rho} B^Y_{\sigma} - \partial_{\sigma} B^Y_{\rho}$ and $F^W_{\rho \sigma}= F^a_{\rho \sigma} T^a = \left(\partial_{\rho} W^a_{\sigma} - \partial_{\sigma} W^a_{\rho} + g \epsilon^{abc} W^b W^c \right) T^a$.\\

For our aim we are interested in the gauge invariant terms which
couple to the Higgs field, in order to reproduce the coupling to the
axion $a_X$ and the SM neutral Golstone boson $a_H$, which in the SM broken phase is :
%\footnote{Note that the interaction terms with the Higgs field disappear in this specific combination.}
\begin{equation}
%\epsilon^{\mu \nu \rho \sigma}  (Z'_{\mu}-\frac{1}{M_{Z'}}\partial_{\mu} a_X)(Z_{\nu}-\frac{1}{M_{Z}}\partial_{\nu} a_H) F^Y_{\rho \sigma}
\epsilon^{\mu \nu \rho \sigma}  {\cal D}_{\mu} \theta_{X}~ {\cal
  D}_{\nu} \theta_{H}~ F^Y_{\rho \sigma} \ ,
\end{equation}
where $\theta_H = a_H/v$. \\

In (\ref{l02}), the coefficient $\delta$  parameterizes the kinetic mixing term of $Z'$
with the hypercharge gauge field. The parameter $\eta$ generates, after
electroweak symmetry breaking, a mass mixing between $Z$ and
$Z'$ that can be estimated to be small in such effective theories and we ignore it in our analysis.
On the other hand,  $b_i$, $c_i$ and $d_i$ in (\ref{lmix}) contain the
possible low-energy three gauge-boson interaction terms. Their values are fixed
by the properties of the more fundamental theory, and in particular
many of them can vanish depending on the spectrum of the sector integrated out.
%We will show in Section \ref{Examples} how different patterns are actually possible. \\
Some remarks are in order : the kinetic mixing term is a general
feature of all the heavy fermion spectra with coupling to both
$U(1)_X$ and $U(1)_Y$. However, it follows from  the Furry's
theorem that the cubic gauge boson interaction terms appear
only when chiral fermions are integrated out. This is then consistent with
the heavy fermions getting masses from Yukawa couplings to the heavy Higgs $S$ breaking $U(1)_X$.
Effective operators like the ones in the second line in
(\ref{lmix}) contain,after electroweak symmetry breaking, the generalized Chern-Simons (GCS) dimension-four operator (\ref{i3})
\begin{equation}
\frac{v^2}{M^2} \ \epsilon^{\mu \nu \rho \sigma} Z'_{\mu} B_{\nu} F_{\rho \sigma}^Y \ . \label{gcs2}
\end{equation}
Since it originates from a dimension six operator, (\ref{gcs2}) is
suppressed by the heavy mass scale $M^2$. As we will see in
Section \ref{UV}, this can also be explicitly checked by using the
formulae for axionic couplings
and GCS terms in \cite{abdk} in the decoupling limit $M \rightarrow
\infty$. As a general result, for one Z' the mass-independent
three gauge-boson interaction terms indeed vanish after imposing the anomaly
cancelations for the heavy spectrum, and the leading non vanishing
contributions come from the dimension-six operators listed previously. It is
interesting to notice that, with the exception of a possible kinetic
mixing in the first line in
(\ref{lmix}), all the operators mixing the (in)visible $Z'$ to the SM
are mass-suppressed and therefore decouple at low-energy, in agreement to the
decoupling theorem \cite{appelquist}.
The only ingredients we need in order to prove this are the SM gauge
invariance in the unbroken phase and CP symmetry of the effective operators in the decoupling limit.\\
Finally, notice that our analysis concerning the effective operators
mixing $Z'$ to the SM gauge bosons, remains valid when the heavy
sector integrated out is free only from mixed gauge anomalies (and for
example with a non-vanishing $U(1)_X^3$ anomaly).
%%%%%%%%%%%%%%%%%%%%%%%%%%%%%%%%%%%%%%%%%%%%%%%%%%%%%%%%%%%%%%%%%%%%%%%%%%%%%%%%%%%%%%%%%%%

Due to the mass suppression, effects of the operators discussed in the
previous paragraph at low-energy are generically suppressed by
$E^2/M^2$ or $v^2/M^2$  and can have important effects only for energies not far below the heavy
fermion masses. For vector-like heavy fermions the suppression is more
severe $E^4/M^4$ due to the charge conjugation invariance
constraints. An obvious question is then if it is possible at all to
generate genuine non-decoupling effects by integrating-out
heavy fermions which are chiral with respect to extra $U(1)$'s but
cancel all triangle gauge anomalies between themselves. For one
massive $U(1)_X$, as we proved above, this is impossible. For two
extra symmetries, $U(1)_X$ and $U(1)_X'$ with Stueckelberg realization
of gauge symmetries,
this is however possible \cite{aw}. Indeed, in this case the dimension-four operator
\begin{equation}
\epsilon^{\mu \nu \rho \sigma} \ {\cal D}_{\mu} \theta_X  \ {\cal D}_{\nu}
\theta_X' \ F_{\rho \sigma}^Y \label{nondecoupled}
\end{equation}
is gauge invariant and can be generated by a heavy chiral but
anomaly-free fermion spectrum charged under the two $U(1)$'s and
simultaneously under the SM. We explicitly verify this in Section 4 by using the formulae of
ref. \cite{abdk}. The term  (\ref{nondecoupled}) provides
an interesting counter-example of the decoupling theorem
\cite{appelquist}, different from the non-decoupling effects studied
in \cite{dhoker}, since the heavy fermionic spectrum, albeit chiral, is free of any gauge anomalies.

%%%%%%%%%%%%%%%%%%%%%%%%%%%%%%%%%%%%%%%%%%%%%%%%%%%%%%%%%%%%%%%%%%%%%%%%%%%%%%%%%%%%%%%%%%%%%%%%%%%%%%%%%%

\section{(In)visible $Z'$ as a mediator of dark matter annihilation}

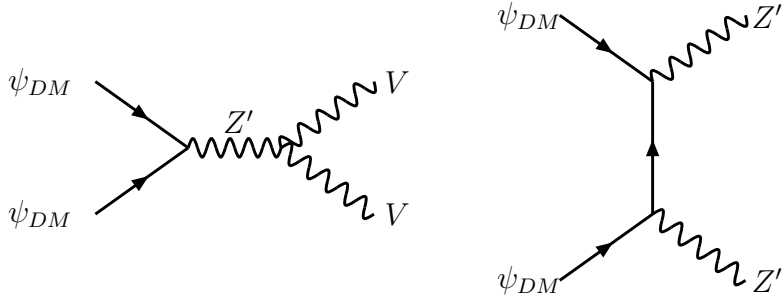
\begin{figure}
\begin{center}
\begin{picture}(100,90)(-30,-5)
\hspace*{-11.5cm}
\SetWidth{1.1}
%%%%%%%%%%%%%%%%%%%%%%%%%%%%%%%%%%%%
%\Text(120,75)[]{\bf a)}
\ArrowLine(205,25)(240,50)
\ArrowLine(205,75)(240,50)
\Photon(240,50)(275,50){3.5}{5}
\Photon(275,50)(310,25){3.5}{5}
\Text(260,60)[]{$Z'$}
\Photon(275,50)(310,75){3.5}{5}
%\Text(240,50)[]{$Z'_{\mu}(p_1)$}
\Text(320,75)[]{$V$}
\Text(320,25)[]{$V$}
\Text(185,75)[]{$\psi_{DM}$}
\Text(185,25)[]{$\psi_{DM}$}
\ArrowLine(380,100)(415,75)
\ArrowLine(380,0)(415,25)
\ArrowLine(415,25)(415,75)
\Photon(415,75)(450,100){3.5}{5}
\Photon(415,25)(450,0){3.5}{5}
%\Text(240,50)[]{$Z'_{\mu}(p_1)$}
\Text(460,100)[]{$Z'$}
\Text(460,0)[]{$Z'$}
\Text(370,100)[]{$\psi_{DM}$}
\Text(370,0)[]{$\psi_{DM}$}
%\Text(270,-30)[]
\end{picture}
 \caption{{\footnotesize Feynman diagrams contributing to the dark matter annihilation.}}
\label{feynannihilation}
\end{center}
\end{figure}

Several Dark Matter candidates have been proposed and widely discussed
in the literature. The distinct phenomenological feature of the present model is a clear dark matter
annihilation signature in the galactic halo.
We are interested in particular in
the trilinear couplings of the form $Z' Z \gamma$ and $Z' Z Z$.
These terms can provide a clear signature for the indirect detection of dark
matter. The main idea is the following: if the dark matter candidate
is lighter than the fermionic sector which we integrated out, the unique tree
level annihilation diagram is given by the exchange of $Z'$.
Then, $Z'$ can couple to the visible sector only via the couplings to
the SM gauge bosons. We also stress that, as shown in
Fig. \ref{feynannihilation}, this could also give one of the very few available
signatures of such (in)visible Z'.
The relevant information for the corresponding analysis is contained
in the operators ${\cal L}_{mix}$ in (\ref{lmix}).
Indeeed, we can easily extract the $Z'VV$ interaction vertices
generated by (\ref{lmix})~:

\begin{eqnarray}
\Gamma^{Z'\gamma Z}_{\mu \nu \rho}(p_3;p_1,p_2)&=&
-8 ~\frac{(d_1-d_2)}{M^2} g_X \sin\theta_W \cos \theta_W
(p_1 + p_2)^\mu \epsilon_{\nu \rho \sigma \tau}p_2^\sigma p_1^\tau
\nonumber
\\
&& - 2 ~ \frac{e \ g_X}{\cos \theta_W \sin \theta_W}\frac{v^2}{M^2}
\left[
c_1 \cos \theta_W + c_2 \sin \theta_W
\right]
\epsilon_{\mu \nu \rho \sigma} p_1^\sigma
\nonumber
\\
\Gamma^{Z'Z Z}_{\mu \nu \rho}(p_3;p_1,p_2)&=&
-4~ \frac{(d_1 \sin^2 \theta_W + d_2 \cos^2\theta_W)}{M^2} g_X
(p_1 + p_2)^\mu \epsilon_{\nu \rho \sigma \tau}p_2^\sigma p_1^\tau
\nonumber
\\
&& - ~\frac{e \ g_X}{\cos \theta_W \sin \theta_W}\frac{v^2}{M^2}
\left[
c_2 \cos \theta_W - c_1 \sin \theta_W
\right]
\epsilon_{\mu \nu \rho \sigma} (p_2^\sigma-p_1^\sigma)
\nonumber
\\
\Gamma^{Z'W^+ W^-}_{\mu \nu \rho}(p_3;p_1,p_2)&=&
-4 ~\frac{d_2}{M^2} g_X
(p_1 + p_2)^\mu \epsilon_{\nu \rho \sigma \tau}p_2^\sigma p_1^\tau
\nonumber
\\
&& -~ \frac{e \ g_X}{\cos \theta_W \sin \theta_W}\frac{v^2}{M^2}c_2
\epsilon_{\mu \nu \rho \sigma} (p_2^\sigma-p_1^\sigma)
\label{vertices}
\end{eqnarray}

\begin{figure}
\begin{center}
\begin{picture}(100,90)(-30,-5)
\hspace*{-11.5cm}
\SetWidth{1.1}
%%%%%%%%%%%%%%%%%%%%%%%%%%%%%%%%%%%%
%\Text(120,75)[]{\bf a)}
\Photon(140,50)(175,50){3.5}{5}
\Photon(175,50)(210,25){3.5}{5}
\Photon(175,50)(210,75){3.5}{5}
\Text(150,63)[]{$Z'_{\mu}(p_3)$}
\Text(230,75)[]{$\gamma_\nu(p_1)$}
\Text(230,25)[]{$Z_\rho(p_2)$}
%\Text(190,0)[]{\footnotesize $\left( - \alpha'_1 \cos \theta_W +   \alpha'_2 \sin \theta_W \right) \epsilon^{\mu \nu \sigma \rho}p_{3\sigma}$}

\Photon(290,50)(325,50){3.5}{5}
\Photon(325,50)(360,25){3.5}{5}
\Photon(325,50)(360,75){3.5}{5}
\Text(300,63)[]{$Z'_{\mu}(p_3)$}
\Text(380,75)[]{$Z_\nu(p_2)$}
\Text(380,25)[]{$Z_\rho(p_3)$}
%\Text(350,0)[]{\footnotesize $ \left( \alpha'_1 \sin \theta_W + \alpha'_2 \cos \theta_W \right) \epsilon^{\mu \nu \sigma \rho} p_{3 \sigma}$}

\Photon(440,50)(475,50){3.5}{5}
\Photon(475,50)(510,25){3.5}{5}
\Photon(475,50)(510,75){3.5}{5}
\Text(450,63)[]{$Z'_{\mu}(p_3)$}
\Text(535,75)[]{$W_\nu^+(p_1)$}
\Text(535,25)[]{$W_\rho^-(p_2)$}
%\Text(500,0)[]{\footnotesize $ \alpha'_2 \cos  \theta_W ~ \epsilon^{\mu \nu \sigma \rho}(p_{3 \sigma}-p_{2 \sigma})$}
\label{feynVVV}
\end{picture}
 \caption{{\footnotesize Three vertices of interest generated by (\ref{lmix}).}}
\end{center}
\end{figure}
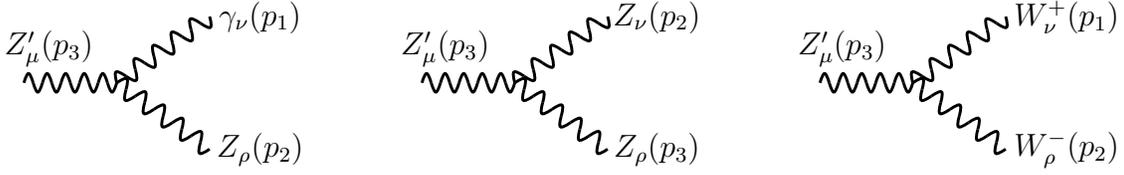

%\begin{figure}
%\begin{center}
%\begin{picture}(100,90)(-30,-5)
%\hspace*{-11.5cm}
%\SetWidth{1.1}
%%%%%%%%%%%%%%%%%%%%%%%%%%%%%%%%%%%%%
%\Text(120,75)[]{\bf a)}
%\Photon(280,50)(315,50){3.5}{5}
%\ArrowLine(315,50)(350,25)
%\ArrowLine(315,50)(350,75)
%\Text(240,50)[]{$Z'_{\mu}(p_1)$}
%\Text(260,50)[]{$Z'_{\mu}$}
%\Text(370,75)[]{$\psi_{DM}$}
%\Text(370,25)[]{$\psi_{DM}$}
%\Text(330,0)[]{\footnotesize $i \frac{1}{4}g' \gamma^{\mu}(V_{DM}-A_{DM}\gamma^5)$}
%\end{picture}
%\caption{\footnotesize{Interaction vertices for $Z'$.}}
%\label{feynDMDM}
%\end{center}
%\end{figure}

\noindent
where $\theta_W$ is the Weinberg angle, $e$ is the electric charge and $M$ is the typical scale in the massive fermionic sector
($\sim$ TeV). As we will see in UV completions discussed in Section 4,
the coefficients ($c_i$, $d_i$) are combinations of gauge charges and
are suppressed by a loop factor
($\sim 10^{-2}$). Before analyzing in more details such a model, it is
interesting to observe the
dependence of the vertices on each coupling.
For instance, $\Gamma^{Z'VV}_{\mu\nu\rho}$ are independent of the coefficients $b_i$ after
symmetrization under the exchange ($p_1 , \nu \leftrightarrow p_2 , \rho$)
and the dependence
of  $\Gamma^{Z'Z\gamma}_{\mu\nu\rho}$ on $d_i$ is proportional to the total momentum
$(p_1+p_2)^\mu$. These facts have
significant consequences on the gamma-ray spectrum. Indeed, it follows
from (\ref{vertices}) that the Z $\gamma$ final state can naturally
compete with the $Z Z$ one in a large part of the parameter space.

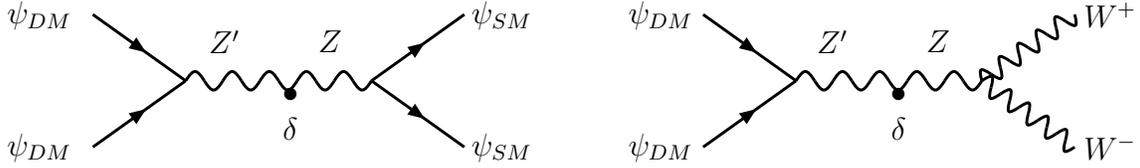
\begin{figure}
\begin{center}
\begin{picture}(100,90)(-30,-5)
\hspace*{-8.5cm}
\SetWidth{1.1}
%%%%%%%%%%%%%%%%%%%%%%%%%%%%%%%%%%%%
%\Text(120,75)[]{\bf a)}

\ArrowLine(75,25)(110,50)
\ArrowLine(75,75)(110,50)
\Photon(110,50)(180,50){3.5}{5}
\ArrowLine(180,50)(215,25)
\ArrowLine(180,50)(215,75)
%\Text(240,50)[]{$Z'_{\mu}(p_1)$}
\Text(125,65)[]{$Z'$}
\Text(150,45)[]{$\bullet$}
\Text(150,32)[]{$\delta$}
%\Line()
\Text(165,65)[]{$Z$}
\Text(230,75)[]{$\psi_{SM}$}
\Text(230,25)[]{$\psi_{SM}$}
\Text(55,75)[]{$\psi_{DM}$}
\Text(55,25)[]{$\psi_{DM}$}

\ArrowLine(305,25)(340,50)
\ArrowLine(305,75)(340,50)
\Photon(340,50)(410,50){3.5}{5}
\Photon(410,50)(445,25){3.5}{5}
\Photon(410,50)(445,75){3.5}{5}
%\Text(240,50)[]{$Z'_{\mu}(p_1)$}
\Text(355,65)[]{$Z'$}
\Text(380,45)[]{$\bullet$}
\Text(380,32)[]{$\delta$}
%\Line()
\Text(395,65)[]{$Z$}
\Text(460,75)[]{$W^+$}
\Text(460,25)[]{$W^-$}
\Text(290,75)[]{$\psi_{DM}$}
\Text(290,25)[]{$\psi_{DM}$}
\end{picture}
 \caption{{\footnotesize R\^ole of the mixing parameter $\delta$.}}
\label{mixing}
\end{center}
\end{figure}

We also notice that all these trilinear coupling can be
written symbolically as $\frac{(\partial)^3}{M^2} V V V$, with $V$ a generic gauge
boson, except for the terms depending explicitly on the Higgs field,
which after electroweak symmetry breaking are of the form
$\frac{v^2}{M^2} \partial V V V$. The ratio between
the two contributions in the
process we are interested in, will then be roughly depending on the
ratio $\frac{v^2}{M_{DM}^2}$. Therefore, the larger the mass of the
dark matter candidate, the smaller the contributions of the terms
related to the operators parameterized by the coefficients $c_i$, with respect to the other ones.\\
Concerning the kinetic mixing $\delta$, its presence induces a
redefinition of the gauge boson mass matrix
eigenvectors and eigenvalues, as extensively studied in the literature
\cite{zprime}. The kinetic mixing also contributes to the DM
annihilation, as shown in Fig. \ref{mixing}. A general comment is that
the smaller the kinetic mixing compared to the $Z'Z \gamma$ vertex,
the cleaner is the monochromatic gamma ray.

We should also parameterize the $\psi^{DM} \psi^{DM} Z'$ coupling. For
a generic DM fermion, the vertex can be written as
%, where $\psi_{DM}$ is the lightest
%stable heavy fermion responsible of the anomaly cancelation process\footnote{
%We have neglected for the rest of the study the $Z Z'$ mixing}.

\begin{equation}
\Gamma_{\psi^{DM} \psi^{DM} Z'} = i \frac{g_X}{4} \gamma^{\mu}(V_{DM} - A_{DM}\gamma^5)
\end{equation}

\noindent
where $V_{DM}$ and $A_{DM}$ are the vectorial and axial couplings, related to the
$U(1)_X$ charges of the Weyl components of $\psi^{DM}$.
For a DM candidate with a pure vectorial coupling, the operators with
coeff. $d_i$ in the third line of eq. (\ref{lmix}) do not contribute
to the DM annihilation amplitudes,  due to the conservation of
the DM vector current. On the other hand, the same operators, in the
case of a DM with a pure axial coupling, contribute to the amplitudes proportionally to the
divergence of the DM axial-vector current and therefore proportionally
to the DM mass $M_{DM}.$
In our study we consider the general case with both vector and axial-vector couplings,
with $(V_{DM},A_{DM})=(1,0)$ for vectorial coupling and $(V_{DM},A_{DM})=(0,1)$
for axial couplings. The specific values of $(V_{DM},A_{DM})$ are not relevant in
our study as they can be absorbed by a redefinition of the $U(1)'$ coupling $g_X$.

\noindent
As noticed above, the important point in the formula (\ref{vertices}) is that,
depending on the relative values
of the coefficients $c_i$ and $d_i$, we obtain different final states in the annihilation
process $\psi^{DM} \psi^{DM} \rightarrow Z' \rightarrow VV$. The most
interesting for us is the $Z \gamma$ final state. Indeed, we can show easily that for such a final state,
because the annihilation occurs for  DM particles at rest,
the energy of the photon is monochromatic\footnote{One should note that models with DM annihilations producing
enhanced gamma ray lines  have also been recently suggested in a completely
different framework in \cite{Gustafsson:2007pc}. } and equal to

\begin{equation}
 E_{\gamma} \ = \  M_{DM} \ \left[ 1- (\frac{M_{Z}}{2M_{DM}})^2
 \right] \ . \label{mono}
\end{equation}

\begin{figure}
    \begin{center}
    \includegraphics[width=3.in]{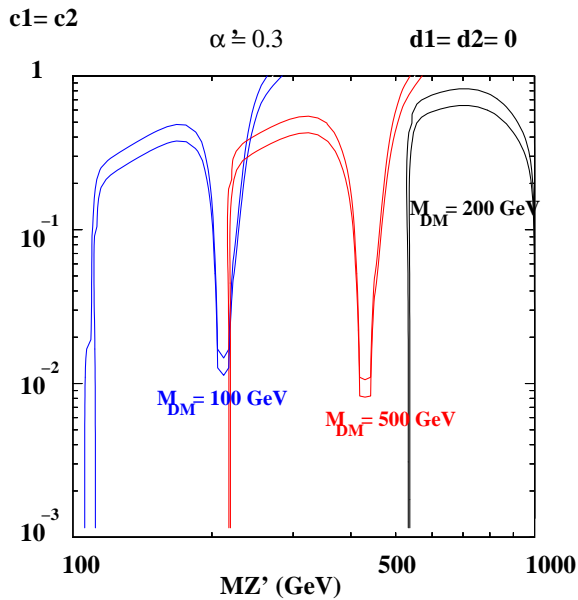}
%    \hspace{1cm}
%    \includegraphics[width=2.22in]{Scanb2001.eps}
          \caption{{\footnotesize
Scan on the mass of $Z'$ (in logarithmic scale) versus the couplings $c_1=c_2$ for
$d_1=d_2=0$ and $M=1$ TeV. We also defined $\alpha'= g_X^2/4 \pi$.
Colored lines represent the WMAP limits on the dark matter relic
density for different values of the dark matter mass.
Notice that the results are invariant under the rescaling
$M\rightarrow \alpha M$, $(c_i,d_i)\rightarrow (\alpha^2 c_i,\alpha^2 d_i)$.
}}
\label{fig:relic}
\end{center}
\end{figure}

\begin{figure}
    \begin{center}
    \includegraphics[width=3.in]{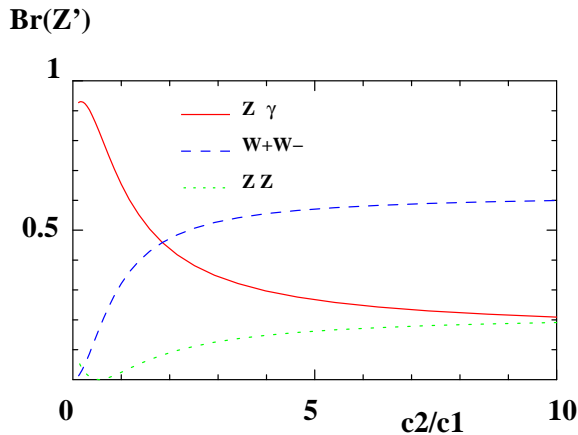}
%    \hspace{1cm}
%    \includegraphics[width=2.22in]{Scanb2001.eps}
          \caption{{\footnotesize
          {\bf
Branching fractions of Z' boson decays into $Z \gamma$ (red full line),
$ZZ$ (green dotted line) and $W^+W^-$ (blue dashed line) as a function
of $c_2/c_1$ for $M_{DM}=200$ GeV, $M_{Z'}=215$ GeV and $d_1=d_2=0$.
}
}}
\label{fig:Br}
\end{center}
\end{figure}

We computed the relic density $\Omega h^2$ using the last released
 version of the Micromegas code
\cite{Belanger:2008sj}, modified to include the (in)visible Z' and its
couplings to the SM.
We show the results in Fig. \ref{fig:relic}  as a scan on ($M_{Z'}$, $c_1=c_2$)
for $d_1=d_2=0$.
The region between the two lines of the same color corresponds to
the 5$\sigma$ region of WMAP \cite{Spergel:2006hy}.
The different dominant annihilation channels contributing to $\Omega h^2$
are depicted in Figs. \ref{feynannihilation}.
We clearly see the p\^ole regions when $M_{Z'} \sim 2 M_{{DM}}$.
In the $Z'$-p\^ole region, the main
annihilation channel for $c_2=d_2 =0$ is the $\gamma Z$ final state at more than 80 \%.
This comes mainly from the reduction of the $Z'ZZ$ coupling  and a smaller phase space
in the final state.
For $c_2, d_2 \not=0$, the annihilation channel
$\psi_{DM} \psi_{DM} \rightarrow Z' \rightarrow W^+ W^-$ becomes
stronger and  $\Omega h^2$ decreases, leading to slight changes in
the Figure  \ref{fig:relic}.
When the
$W^+W^-$ channel is open ($d_2, c_2 > d_1, c_1$), this final state is the dominant one
at 60\%.
We plot in Fig. \ref{fig:Br} the branching fraction of $Z'$ into the $Z \gamma$
(red full line), $ZZ$ (green dotted line) and $W^+W^-$ (blue dashed line) as a function
of $c_2/c_1$ for $M_{DM}=200$ GeV and $M_{Z'}=215$ GeV.
$W^+W^-$ final state begins to be dominant for $c_2 \sim 2 c_1$.
Another interesting feature in regions where the $Z'$ is light
($M_{Z'}< M_{{DM}}$) is that we observe a new
zone, respecting WMAP, almost independent of the value of $d_1$. This region comes
from the t-channel annihilation into $Z'Z'$ final state, depicted in
Fig. \ref{feynannihilation}.
Depending on the energy of the $Z'$ in the final state, we can observe its decay modes to
SM gauge bosons.

\begin{figure}
    \begin{center}
    {\bf a)}
    \includegraphics[width=3.in]{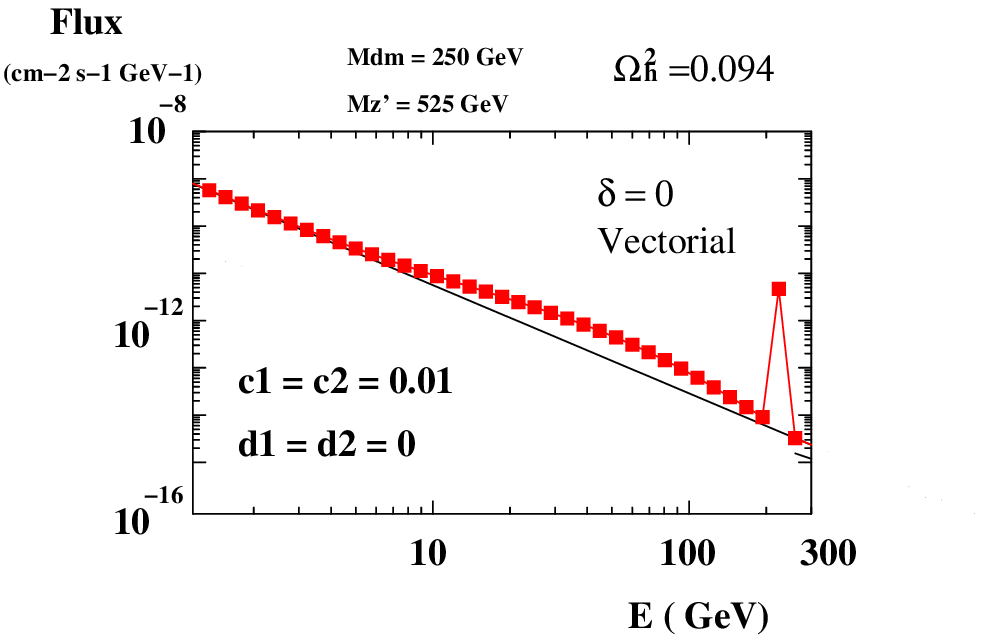}
    {\bf b)}
    \includegraphics[width=3.in]{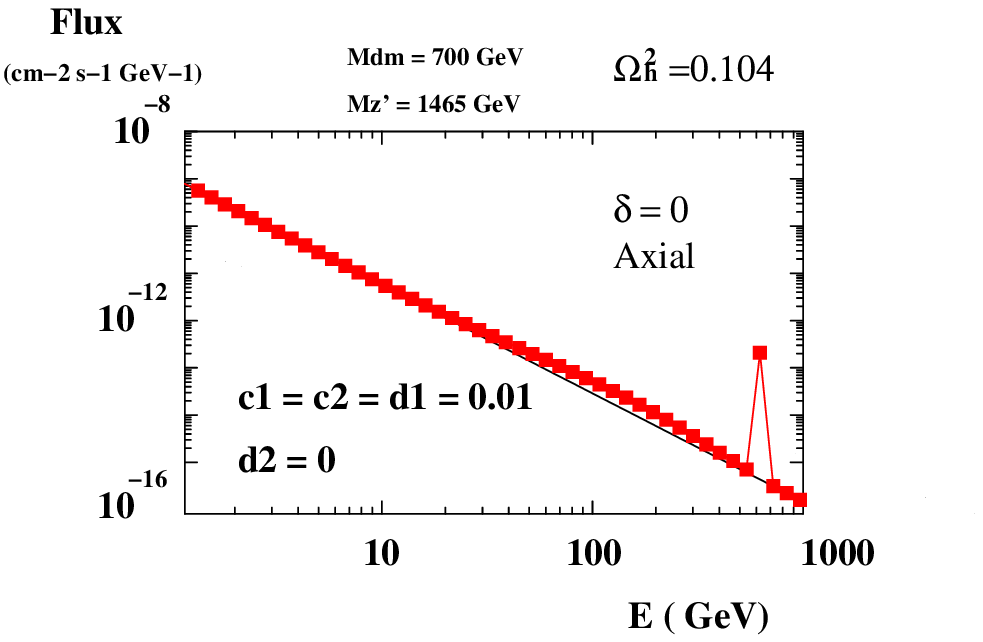}
    {\bf c)}
    \includegraphics[width=3.in]{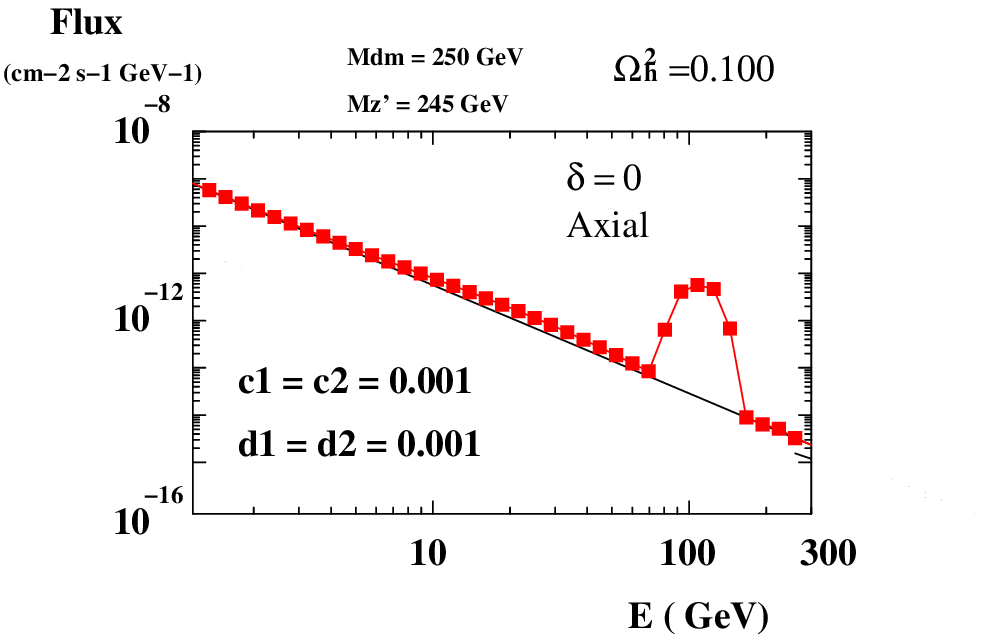}
    {\bf d)}
    \includegraphics[width=3.in]{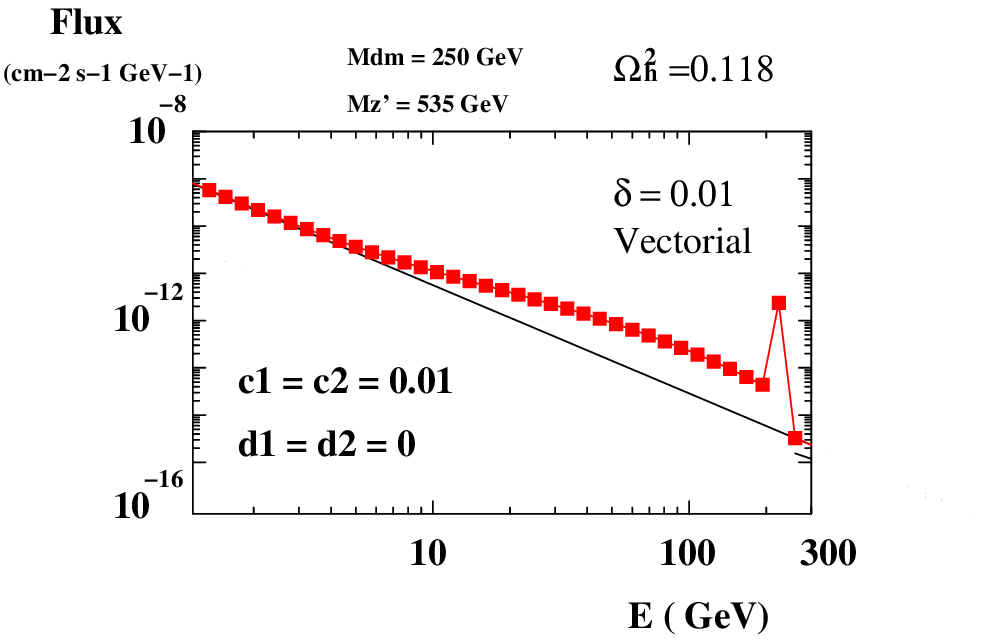}
              \caption{{\footnotesize
Typical example of a gamma-ray differential spectrum for different masses of dark matter
and $Z'$ and $Z-Z'$ mixing angle, compared with the background (black line \cite{Bernal:2008zk}).
All fluxes are calculated for a classical NFW halo profile and $M=1$ TeV.
"Axial" and "vectorial" stands for the nature of the $\psi_{DM}\psi_{DM}Z'$ coupling.}
}
\label{fig:spectrum}
\end{center}
\end{figure}

Concerning the direct detection prospects, it will be hard to see any spin--dependent
or spin--independent signal. Indeed, the idea of direct detection experiment is based upon
the measurement of the recoil energy of a target nucleus hit by a dark matter particle.
 As no effective coupling exists between the $Z'$ and the constituent quarks of the proton,
 the cross section of such process is simply suppressed\footnote{This is one of the main reason that the (in)visible $Z'$
 will be hardly visible at LHC. Indeed, the relevant Feynman diagrams are the same, whereas the main production channel
 of the $Z'$ is by vector-vector fusion, as described in \cite{wells}.}.
 However the indirect detection possibilities seem more promising
 as the dark matter annihilate only into $WW$, $ZZ$ or $Z\gamma$ final states.
 The monochromatic photon present in
 the $Z\gamma$ final state could be a smoking gun signal of such
 models. Indeed, monochromatic processes exist also in supersymmetric
 or KK-like models at one loop order, but they are invisible after including the
 total amount of diffuse flux coming from the tree level processes.
 In the case of an (in)visible $Z'$, all the final states come with an amplitude of the
 same order of magnitude : the $\gamma$-monochromatic line can easily be disentangled from
 the diffuse background. We plotted in Figs. \ref{fig:spectrum} the diffuse gamma
 fluxes from the galactic center that we expect in different scenarios for a DM mass of 250 GeV
 and 700 GeV.  For heavy axial dark matter, (Fig.\ref{fig:spectrum}b) more
than 90\% of the signal is coming from the term proportional to $d_i$.
 We used the Pythia Monte Carlo to simulate the gamma-ray spectrum using an analysis similar to that performed in \cite{Bernal:2008zk}.

 For a $Z'$ mass close to pole $M_{Z'} \sim 2 M_{DM}$, the main process contributing
 to the relic abundance is the s-channel exchange  of a $Z'$: the monochromatic line (\ref{mono}) is clearly
 visible (Fig. \ref{fig:spectrum}a and b).
 However, if the $Z'$ is lighter than $\psi_{DM}$, the t-channel annihilation
  $\psi_{DM} \psi_{DM} \rightarrow Z' Z'$ dominate the relic density annihilation processes.
  The $Z'$ decays finally into $ZZ$ or $Z\gamma$, but not exactly at rest : we measure the "would-be"
  monochromatic $\gamma$-ray line, deformed by the kinetic component of the $Z'$
  (Fig. \ref{fig:spectrum}c). In some cases, the result can be spectacular
 and could be seen by the satellite GLAST/FERMI-LAT \cite{Baltz:2008wd} after 5 years of data taking\footnote{Work in progress.}.
 Finally, we also checked how the kinetic mixing between $Z$ and $Z'$
 could play an important role in such analysis.
 Even with large mixing (Fig. \ref{fig:spectrum}d), we are still able to observe a (slightly reduced)
 gamma-ray line of  the same order of magnitude as the continuous signal. The reduction of the amplitude
 comes from the fact that the diagrams depicted
 in Fig. \ref{mixing} also contribute to the annihilation of the DM particle . The chosen value of the mixing $\delta $ is the
 maximal value consistent with the direct detection limit coming from the proton--dark matter elastic scattering cross section
 ($\sigma_{\Psi^{DM}-p}^{SI}\sim 10^{-8}$pb) \cite{Angle:2007uj}.
Let us mention that recent works (\cite{Feldman:2007wj} and
 references therein) studied Stueckelberg $Z'$ extensions
 with Z-Z' kinetic mixing, but without the Chern-Simons terms.

In the case of two $Z'$, as discussed in Section 2 and further
discussed in Section 4, there is a genuine non-decoupling
operator (\ref{nondecoupled}). If its coefficient is
comparable or dominant over the kinetic mixings $Z Z'$ and $Z Z''$, the dark-matter annihilation proceeds via
the process
$\psi^{DM} \psi^{DM} \ \rightarrow \ {\rm virtual} \ Z' \ \rightarrow \ Z''\ \gamma$ ,
which is unsuppressed by the heavy mass and generates a clean gamma ray signal at an energy
$ E_{\gamma} \ = \  M_{DM} \ \left[ 1- (\frac{M_{Z''}}{2M_{DM}})^2
 \right]$ .

\section{ UV renormalizable theories } \label{UV}

Finally, we would like to discuss possible UV completions that give
the earlier discussed effective operators.
The question  if such patterns  emerge in
renormalizable quantum field theories was addressed in \cite{abdk}.
The framework is a consistent ({\it i.e.} anomaly-free) and
renormalizable gauge theory with spontaneously-broken gauge symmetry
via the Brout-Englert-Higgs mechanism. Through appropriate Yukawa
couplings, some large masses can be
given to a subset of the fermions . We consider the general case of several spontaneously broken
$U(1)$'s. We denote by $\psi^{(h)}_{L,R}$ such massive chiral fermions.
Their $U(1)_i$ charges are $X^{(h)i}_{L,R}$. In the sequel, we briefly
review and adapt the results of the calculation \cite{abdk} of the effective
axion and GCS couplings at low-energy , generated by the loops of  the heavy
chiral fermions. In what follows, we write explicitly only the
massive gauge fields $A_i$ and we consider for simplicity a number of
Higgs fields $S_i$ equal to the number of massive $U(1)$'s. This allows to simplify the formulae, but the results can be easily extended to a more general case. Therefore we can assume that the Higgs field $S_{{i}}$ has charge 1 under the gauge transformation related to the vector field $A_{{i}}$ (more complicated charge assignments can be reduced to this one after rotations and redefinitions).

The relevant terms in the effective action of the heavy fermion sector of the theory are
\bea L_h \
&=& \ {\bar \psi}^{(h)}_L \left( i \gamma^{\m}
\partial_{\m} + g^i X^{(h)i}_L \gamma^{\m} A_{\mu}^i \right)
\psi^{(h)}_L + {\bar \psi}^{(h)}_R \left( i \gamma^{\m}
\partial_{\m} + g^i X^{(h)i}_R \gamma^{\m} A_{\mu}^i \right)
\psi^{(h)}_R
\nonumber \\
&& - \left( {\bar \psi}^{(h)}_L M^{(h)} \psi^{(h)}_R
+ {\rm h.c.} \right) \ , \label{decoupling1} \eea
where $M^h$ is the mass matrix of heavy fermions, with matrix elements
\begin{eqnarray}
&& M^{(h)}_{ab} \ = \ \lambda_{ab}^h  \ S_{i} \qquad {\rm case \ (a) }
\quad {\rm or}  \nonumber  \\
&& M^{(h)}_{ab}
 \ = \  \lambda_{ab}^h  \  {\bar S}_{i} \qquad {\rm case \ (b) } \ , \label{decoupling01}
\end{eqnarray}
where  $S_{i}$ is the Higgs field  of charge $+1$ under the gauge group
$U(1)_{i}$ and singlet with respect to the other gauge groups.
The Higgses spontaneously break the abelian gauge symmetries via their vevs, $\langle S_i
\rangle = V_i$. For every gauge group $U(1)_i$ , there is a set $h_i$ of fermions We charges of the fermions satisfy the relations
\be X^{(h_i)i}_{L} \ - \ X^{(h_i)i}_{R} \ = \ \pm 1 \ \equiv \ \epsilon^{(h_i)i} \
% \not= \ 0
\label{decoupling2} \ee
in order to couple with the Higgs field $S_i$.
The fermions we are considering are charged both under the SM gauge
group and the additional $U(1)$'s. Whereas they are {\it chiral} wrt
the $Z'$-type symmetries, they are {\it vector-like} wrt the SM gauge
group, as required by the existence of the mass terms in (\ref{decoupling01}).
If the associated Yukawa coupling eigenvalues are
large,  $\lambda_{ab}^h \gg g_i$, spontaneous symmetry breaking
generates large Dirac fermion masses. We consider the heavy
fermion decoupling limit, with fixed Higgs vev's and fixed gauge
boson masses, whereas $M^{(h)} \rightarrow \infty$.  Since we are
interested in the (in)visible $Z'$ where the SM fermions are neutral
under the massive $U(1)$'s, the heavy fermion sector is by itself anomaly-free
\be \sum_h (X_L^i X_L^j X_L^k - X_R^i
X_R^j X_R^k )^{(h)} \ = \ 0 \ . \label{decoupling02} \ee
We are interested in low-energy couplings generated by the loops of the
heavy fermions. After gauge symmetry breaking, we parameterize the scalar
fields  by
\be S_i \ = \ (V_i + s_i) \ e^{\frac{i \ a_i}{ V_i}} \
, \label{decoupling3} \ee
where $s_i$ are massive Higgs-like
fields and $a_i$ are axions.
The gauge transformations of gauge fields and axions are
\be
\delta
A_{\mu}^i \ = \
\partial_{\mu} \alpha^i \quad , \quad \delta a_i \ = \ V_i \  \
\alpha^i \ .\label{decoupling4}
\ee
%
%where $X^i_I$ are the $U(1)_i$ charges of $S_I$.

The GCS terms and axionic couplings can be computed by performing a
diagrammatic computation with the action
(\ref{decoupling1}), by starting from the corresponding three gauge
boson amplitude induced by triangle diagram loops of heavy fermions
and expanding in powers of external momenta $k / M^{(h)}$. We define
the effective action after integrating out the heavy states by
\bea {\cal S} & = & - \sum_i  \int {1 \over 4}
F_{i,\m\n}  F_i^{\m\n} + {1 \over 2} \int   \sum_i ( \partial_\m a^i - g_i
V_i A_{\m}^i)^2
\ , \nonumber \\
&& + {1 \over 96 \pi^2} C^{i}_{ij}~\epsilon^{\mu \nu \rho \sigma} \int
 a^i F^i_{\mu \nu} F^j_{\rho \sigma} + {1 \over
48 \pi^2} E_{ij,k}~ \epsilon^{\mu \nu \rho \sigma} \int A^i_{\mu} A^j_{\nu} F^k_{\rho \sigma} \ ,
\label{decoupling5}
\eea
where the coefficients $E_{ij,k}$ satisfy the cyclic relation
\begin{equation}
 E_{ij,k} \ + \ E_{jk,i} \ + \  E_{ki,j} \ =  \ 0 \label{cocycle}
\end{equation}
and the gauge invariance conditions, in the presence of an anomaly free spectrum, read
\begin{eqnarray}
&& C_{jk}^{i} g_i V_i \ - \ E_{ij,k} \ - \ E_{ik,j} \ = \ 0 \ , \nonumber \\
&& C_{jk}^{i} g_i V_i \ + \ C_{ki}^{j} g_j V_j \ + \ C_{ij}^{k} g_k
V_k \ = \ 0 \ . \label{cond}
\end{eqnarray}
One can easily find the solution of (\ref{cond})
\begin{equation}
E_{ij,k} \ = \ \frac{1}{3} \left( g_i V_i C_{jk}^i \ - \ g_j V_j C_{ik}^j\right) \ .
\end{equation}
The result obtained in \cite{abdk}, in the decoupling limit $M^{(h)}
\rightarrow \infty$ with finite Higgs vev's $V_i $ is
\bea
&& E_{ij,k} \ = \ {1 \over 4}
 \sum_h (X_L^i X_R^j- X_R^i X_L^j)^{(h)} (X_R^k+ X_L^k)^{(h)} \ , \nonumber \\
&& C^I_{ij} \ = \  {1 \over 4 g_I V_I} \ \sum_{h_I} \epsilon^{(h_I)I} [ 2 (X_L^i X_L^j + X_R^i
X_R^j) +  X_L^i X_R^j + X_R^i X_L^j ]^{(h_I)} \ ,
%\forall h_I \ ,
%&& M^I_i \ = \ V_I \ X_I^i \ = \ V_I \ (X_L^i-X_R^i)^{(h_I)} \quad ,
%\quad {\rm for \ every \ }
\label{decoupling6}
\eea
where the index $h_I$ in (\ref{decoupling6}) refers to the heavy
fermionic spectrum coupling to the axion $a_I$.
%In addition, we
%introduced for convenience of notation the signs
%\begin{equation}
%\epsilon_h = + 1 \quad {\rm for \ case \ (a) \ in \ (\ref{decoupling01})} \quad ,
%\quad   \epsilon_h = - 1 \quad {\rm for \ case \ (b) \ in \ (\ref{decoupling01})} \
%. \label{decoupling7}
%\end{equation}
Notice that, while within the more general framework of \cite{abdk}
the GCS terms $E_{ij,k}$ had a freedom in their definition related to
the different possible distribution of the gauge anomalies in the
low-energy theory among different $U(1)$ currents, in the context of
the present paper, they are uniquely fixed since the low-energy
spectrum is neutral under the massive abelian gauge fields.

Actually, due to gauge invariance of the low-energy effective action,
the unique gauge invariance combination of GCS and axionic terms is \cite{ferrara,abdk,aw}
\begin{equation}
\frac{1}{48 \pi^2} d_{ij,k} \ \epsilon^{\mu \nu \rho \sigma} \ (\partial a^i - g^{i} V^{i} A^i)_{\mu} \ (\partial a^j - g^{j} V^j A^j)_{\nu} \ F^k_{\rho \sigma} \ , \label{decoupling8}
\end{equation}
which justifies the form of the operators used in Section 2, in terms
of the Stueckelberg gauge-invariant combinations.
They lead to the couplings
\begin{equation}
E_{ij,k} \ = \ d_{ij,k} g^{i} g^{j} V^{i} V^{j}  \quad , \quad
C^i_{jk} \ = \ d_{ij,k} \ g^j V^j + d_{ik,j} g_k V_k \ . \label{decoupling9}
\end{equation}

We now analyze in a model-independent way the resulting low-energy
couplings in the cases of one and two extra $U(1)$'s, using (\ref{decoupling6}) and the anomaly
cancelation constraints, and compare the results with the operatorial analysis performed in
Section 2. We don't need to assume that the
whole mass matrix comes from the breaking of the additional $U(1)$'s. For appropriate
quantum numbers, some SM like mass entries
$\lambda'_{ab} H   {\bar \psi}^{(h,a)}_L  \psi^{(h,b)}_R $,
where $H$ is the SM Higgs, can also exist without changing our
conclusions, as long as we keep $\lambda'_{ab} \langle H \rangle $ fixed
in the decoupling limit $M^{(h)} \rightarrow \infty$.
If the DM is chiral under Z' and a singlet under the SM , we need of course to assume
that there exists a heavier partner singlet under the SM which cancel its $U(1)_{Z'}^3$ anomaly.

%%%%%%%%%%%%%%%%%%%%%%%%%%%%%%%%%%%%%%%%%%%%%%%%%%%%%%%%%%%%%%%%%%%%%%%%%%%%%%%%%%%%%%%%%%%%%%%%%%%%%%
\subsection{One $Z'$}

We are interested in the $XYY$ GCS term, where $X \equiv Z'$ is the
massive gauge boson and $Y$ is the hypercharge one. The relevant information in
the high-energy spectrum is encoded in the  $Y$ and $X$ charges
\begin{eqnarray}
&& \qquad \qquad Y \qquad \qquad X \nonumber \\
&& \Psi_L^a \qquad \ \ y_a \qquad \qquad x_a \nonumber \\
&& \Psi_R^a \qquad \ \ y_a \qquad \qquad x_a - \e_a \label{onez}
\end{eqnarray}
We denote by $l_a = dim R_a$ the dimension of each fermion
representation. The mixed anomaly $Tr (Y^2 X)$, the GCS
coeff. $E_{XY,Y}$ and the axionic\footnote{In this case, there is only one
axion.} coupling $C_{YY}$ are computed to be
\begin{eqnarray}
&& Tr \ (Y^2 X) \ = \ \sum_a l_a \e_a y_a^2 \quad ,\quad
E_{XY,Y}  \ =  \ \frac{1}{2} \sum_a l_a \e_a y_a^2 \ , \nonumber \\
&& C_{YY}  \ = \ \frac{3}{2 g V} \ \sum_a l_a \e_a y_a^2 \ . \label{decoupling10}
\end{eqnarray}
Since they are all proportional to each other and we consider
anomaly-free spectra  $Tr \ (Y^2 X)=0$, the GCS and axionic couplings
vanish in the decoupling limit, in agreement with the effective
operator analysis performed in section \ref{section1}.  On the other
hand, the dimension-six operators in (\ref{lmix}) certainly do exist and
are the ones we took into account in the phenomenological analysis performed
in Section 3.
%%%%%%%%%%%%%%%%%%%%%%%%%%%%%%%%%%%%%%%%%%%%%%%%%%%%%%%%%%%%%%%%%%%%%%%%%%%%%%%%%%%%%%%%%%%%%%%%%%%%%%%
\subsection{Two $Z'$}

In contrast with the case of one $Z'$, the effective operator analysis in
%\ref{subsection2}
Section 2	revealed the
possible existence of the non-decoupling dimension-four operator
(\ref{nondecoupled}). In order to check its existence in the
UV theory with heavy chiral fermions,  we consider the following table
of charge assignements
\begin{eqnarray}
&& \qquad \qquad Y \qquad \qquad X_1 \qquad \qquad X_2 \nonumber \\
&& \Psi_L^a \qquad \ \ y_a \qquad \qquad x_a \qquad \qquad \ z_a \nonumber \\
&& \Psi_R^a \qquad \ \ y_a \qquad \qquad x_a - \e_a \qquad \ z_a \nonumber \\
&& \chi_L^{m} \qquad \ \ y_m \qquad \qquad x_m \qquad \qquad z_m \nonumber \\
&& \chi_R^m \qquad \ \ y_m \qquad \qquad x_m \qquad \qquad z_m  - \e_m
\label{twoz}
\end{eqnarray}
As transparent from the table, the first group of fermions $\Psi$ acquire
masses from the first Higgs field breaking $X_1$, whereas the second group of fermions $\chi$ acquire
masses from the second Higgs field breaking $X_2$.
Analogously to the previous example, we define $l_m = dim R_m$. In
this case we are interested in $E_{X_1X_2,Y}$ and the two axionic
couplings $C_{X_2Y}^{1}$ and $C_{X_1Y}^{2}$. Using again
(\ref{decoupling6}), we find

\begin{eqnarray}
&& Tr \ (X_1 X_2 Y) \ = \ \sum_a l_a \e_a y_a z_a + \sum_m l_m
\e_m x_m y_m   \ , \nonumber \\
&& E_{X_1X_2,Y}  \ =  \frac{1}{2} \left( \sum_a l_a \e_a y_a z_a - \sum_m l_m \e_m x_m y_m \right) \ ,  \nonumber \\
&& C_{X_2Y}^{X_1}  \ = \ \frac{3}{2 g_1 V_1} \ \sum_a l_a \e_a y_a z_a \quad , \quad
C_{X_1Y}^{X_2}  \ = \ \frac{3}{2 g_2 V_2} \ \sum_m l_m \e_a x_m y_m  \ .
 \label{decoupling11}
\end{eqnarray}
In this case, by imposing cancelation of the mixed anomaly
$Tr \ (X_1 X_2 Y)=0$, we find that the GCS and the two axionic couplings
precisely fit into the gauge invariant term
\begin{equation}
 d_{X_1X_2,Y}  \ \epsilon^{\mu \nu \rho \sigma} \ (\partial a_1 - g_1 V_1 X_1)_{\mu}  ( \partial a_2- g_2 V_2 X_2)_{\nu} \ F^Y_{\rho \sigma} \ , \label{decoupling12}
\end{equation}
with $d_{X_1X_2,Y} = \frac{1}{g_1 g_2 V_1 V_2} E_{X_1 X_2, Y} $, which is the non-decoupling operator (\ref{nondecoupled}) we were searching for. It is also easy to compute $E_{Y X_1,X_2}$ and $E_{X_2 Y,X_1}$ and check (\ref{cocycle}) and (\ref{cond}).
There are other gauge invariant operators purely within
the $Z'$ sector that are however irrelevant for our purposes.

%%%%%%%%%%%%%%%%%%%%%%%%%%%%%%%%%%%%%%%%%%%%%%%%%%%%%%%%%%%%%%%%%%%%%%%%%%%%%%%%%%%%%

\section{Conclusions}

In this paper we studied  the consequences of an extension of the standard model containing
an invisible extra gauge group under which the SM particles are neutral. We showed
that effective operators mixing the two sectors are generated by loops of heavy
fermions, which are chiral wrt U(1) and are vector-like wrt to the SM gauge group. This implies in particular that
the decoupling limit is taken by considering large SM invariant masses $M >> v$ of the heavy fermions, whereas keeping fix SM
like masses $m \sim v$. The induced operators mix SM with $Z'$ via a $Z' V V $ vertices. If the lightest fermion in the $Z'$ sector is stable, the induced operators allow for its annihilations that can give rise to a viable dark
matter candidate. Its annihilations produce clean visible signals through a gamma-ray line
 which seems to be quite an universal clear feature of this type of
constructions. The fact that only one gamma-ray line is produced  and the fact that no signal is expected from direct
detection experiments can be a distinctive signature of the model. Indeed a
supersymmetric neutralino or inert Higgs scalar for instance would instead annihilate into
$Z\gamma$ $and$ $\gamma \gamma$ final states with similar ratios
and interact non-trivially with the nuclei.
Such smoking gun signatures could be observable by near future experiments like
FERMI/GLAST or,  after more data taking,  by the HESS/MAGIC telescopes.

From a theoretical side, we showed that heavy, chiral wrt $U(1)$ but
vector-like wrt the SM gauge group, fermions can generate the
effective operators (\ref{lmix}), even if they cancel among themselves all gauge
anomalies. Since we discussed only the case where the heavy fermion masses
which decouple are SM invariant, in the decoupling limit
the effective operators do respect the SM gauge symmetry. This is to be contrasted with the case of decoupling a SM-like mass
\cite{dhoker}, where the low-energy, non-decoupling effective operators arise only in the broken phase, after electroweak symmetry
breaking.  For the same reason, the effective operators realize the $Z'$ gauge symmetry in the broken phase.

As a result, in the case of one $Z'$, in addition to the well-studied $Z'-Z$ kinetic mixing and $Z'-Z$ mass mixing described in
eq. (\ref{l02}), the other operators mixing SM with $Z'$ are the dimension six operators, eqs. (\ref{lmix}). Although they contain
a heavy-mass suppression, they can induce the viable dark-matter annihilation channel discussed in detail in Section 3. On the other
hand, for two $Z'$, as discussed briefly in Section 2 and further in Section 4, there is a genuine non-decoupling producing an
unsuppressed process  $\Psi^{DM} \Psi^{DM} \ \rightarrow \ {\rm virtual} \ Z' \ \rightarrow \ Z'' \ \gamma$, if both additional
gauge bosons are light, in particular if $M_{Z''} < 2 M_{DM}$.
The class of (in)visible $Z'$ discussed in our paper is another
example of an  "abelian hidden sector" \cite{ringwald}, with potential
collider signatures awaiting a dedicated analysis.
It would be interesting to perform a systematic study of the effects
of the effective operators (\ref{lmix}) at low-energy from a
decoupling perspective. It is interesting to note that a recent study \cite{Me}
in the context of the Green-Schwarz mechanism gave similar phenomenological features.

%%%%%%%%%%%%%%%%%%%%%%%%%%%%%%%%%%%%%%%%%%%%%%%%%%%%%%%%%%%%%%%%%%%%%%%%%%%%%%%%%%%%%%%%%%%%%%%
\section*{Acknowledgments}{ Work partially supported by
the European ERC Advanced Grant 226371 MassTeV, by the CNRS
PICS no. 3747 and 4172, in part by the grant ANR-05-BLAN-0079-02, in part by the RTN contracts MRTN-CT-2004-005104 and MRTN-CT-2004-503369,
and the European contract MTKD-CT-2005-029466. E.D. and Y.M. would like to thank the
Institute for Theoretical Physics of Warsaw form warm hospitality and
financial support via the "Marie Curie Host Fellowship for Transfer of
Knowledge" MTKD-CT-2005-029466. The work of Y.M. is partially supported by the PAI programm PICASSO under contract
PAI--10825VF. He would like to thank the European Network of
Theoretical Astroparticle Physics ILIAS/N6 under contract number
RII3-CT-2004-506222 and the French ANR project PHYS@COLCOS for
financial support. The work of A.R. was  supported
by the European Commission Marie Curie Intra-European
Fellowships under the contract N 041443. The authors are grateful to
G. Belanger and S. Pukhov for substantial help with Micromegas and M. Berg and A. Hebecker for useful discussions.}

\nocite{}
\bibliography{bmn}
\bibliographystyle{unsrt}

\end{document}